# CSR INDUCED MICROBUNCHING GAIN ESTIMATION INCLUDING TRANSIENT EFFECTS IN TRANSPORT AND RECIRCULATION ARCS*

C. -Y. Tsai#, Department of Physics, Virginia Tech, VA 24061, USA
D. Douglas, R. Li, and C. Tennant, Jefferson Lab, Newport News, VA 23606, USA

*Abstract*

The coherent synchrotron radiation (CSR) of a high brightness electron beam traversing a series of dipoles, such as transport or recirculation arcs, may result in the microbunching instability (μBI). To accurately quantify the direct consequence of this effect, we further extend our previously developed semi-analytical Vlasov solver [1] to include more relevant coherent radiation models than the steady-state free-space CSR impedance, such as the entrance and exit transient effects derived from upstream beam entering to and exiting from individual dipoles. The resultant microbunching gain functions and spectra for our example lattices are presented and compared with particle tracking simulation. Some underlying physics with inclusion of these effects are also discussed.

## FREE-SPACE CSR IMPEDANCES

For an ultrarelativistic beam traversing an individual dipole, the steady-state CSR impedance in free space can be expressed as [2,3]

$$Z_{CSR}^{ss}(k(s);s)=\frac{-ik(s)^{1/3}A}{|\rho(s)|^{2/3}} \quad (1)$$

where $k = 2\pi/\lambda$ is the modulation wave number, $\rho$ is the bending radius, and the constant $A \approx -0.94 + 1.63i$.

Prior to reaching steady-state interaction, the beam entering a bend from a straight section would experience the so-called entrance transient state, where the impedance can be obtained by Laplace transformation of the corresponding wakefield [4,5]

$$Z_{CSR}^{ent}(k(s);s)=\frac{-4}{s^*}e^{-4i\mu(s)}+\frac{4}{3s^*}(i\mu(s))^{1/3}\Gamma\left(\frac{-1}{3},i\mu(s)\right) \quad (2)$$

where $\mu(s) = k(s)z_L(s)$, $s^*$ the longitudinal coordinate measured from dipole entrance, $z_L=(s^*)^3/24\rho^2$ and $\Gamma$ the upper incomplete Gamma function.

In addition, there are exit CSR transient effects as the beam exits from a dipole. For the case with fields generated from an upstream electron (at retarded time) propagating across the dipole to downstream straight section, i.e. Case C of Ref. [6], the corresponding impedance can be similarly obtained by Laplace transformation:

$$Z_{CSR}^{exit}(k(s);s)=\frac{-4}{L_b+2s^*}e^{\frac{-ik(s)L_b^2}{6|\rho(s)|^2}\left(L_b+3s^*\right)} \quad (3)$$

where $s^*$ is the longitudinal coordinate measured from dipole exit and $L_b$ is the dipole length.

Instead of performing Laplace transformation of the wakefield expression for the case with fields generated from an electron (at retarded time) within a dipole propagating downstream the straight section, here we use Bosch's expression [7] for the exit transient impedance:

$$Z_{CSR}^{drif}(k(s);s)\approx\begin{cases}\frac{2}{s^*}, \text{ if } \rho^{2/3}\lambda^{1/3}\le s^*\le\lambda\gamma^2/2\pi\\ \frac{2k(s)}{\gamma^2}, \text{ if } s^*\ge\lambda\gamma^2/2\pi\\ 0, \text{ if } s^*<\rho^{2/3}\lambda^{1/3}\end{cases} \quad (4)$$

where $s^*$ is the longitudinal coordinate measured from dipole exit. This expression assumes the exit impedance comes primarily from coherent edge radiation in the near-field region (i.e. $z < \lambda\gamma^2$), and in our simulation we only include such transient effects [Eq. (3) and (4)] right after a nearest upstream bend. Here we note that these CSR models are valid only when the wall shielding effect is negligible. The wall shielding effect becomes important when the distance from the beam orbit to the walls $h$ is to satisfy $h\le(\rho\lambda^2)^{1/3}$. We further note that the above impedance models, Eqs. (1-3), assume the beam is at ultrarelativistic energy. It is interesting to examine how the above models deviate for non-ultrarelativistic beams, and this is currently under study in parallel [8].

## NUMERICAL METHODS

To quantify the μBI in a single-pass system, we would estimate the microbunching amplification factor $G$ (or, gain) by two methods. One is to solve the linearized Vlasov equation [9] using given impedance models [e.g. Eqs. (1-4)]. The other, served as a benchmarking tool, is by ELEGANT tracking [10,11]. For the former, we actually solve the general form of Volterra integral equation [9,12]

$$g_k(s)=g_k^{(0)}(s)+\int_0^s K(s,s')g_k(s')ds' \quad (5)$$

where the kernel function is particularly expressed as

$$K(s,s')=\frac{ik}{\gamma}\frac{I(s)}{I_A}C(s')R_{56}(s'\to s)Z(kC(s'),s')\times[\text{Landau damping}] \quad (6)$$

for the [Landau damping] term

$$[\text{Landau damping}]=\exp\left\{\frac{-k^2}{2}\left[\varepsilon_{x0}\left(\beta_{x0}R_{51}^2(s,s')+\frac{R_{52}^2(s,s')}{\beta_{x0}}\right)+\sigma_\delta^2R_{56}^2(s,s')\right]\right\} \quad (7)$$

with $R_{56}(s'\to s)=R_{56}(s)-R_{56}(s')+R_{51}(s')R_{52}(s)-R_{51}(s)R_{52}(s')$ and $R_{5i}(s,s')=C(s)R_{5i}(s)-C(s')R_{5i}(s')$. Here the kernel function $K(s,s')$ describes CSR effect, $g_k(s)$ the resultant bunching factor as a function of the longitudinal position

* This material is based upon work supported by the U.S. Department of Energy, Office of Science, Office of Nuclear Physics under contract DE-AC05-06OR23177.
#jcytsai@vt.edu

given the wavenumber $k$, and $g_k^{(0)}(s)$ is the bunching factor without CSR perturbation.

Here we have made the coasting beam approximation, i.e. the modulation wavelength is much shorter compared with the whole bunch duration. The transport functions $R_{5i}(s)$ ($i$ = 1, 2, 3, 4, 6) are adopted from ELEGANT with slight modification to account for "non-ultrarelativistic" contribution $\left[R_{56}(s) \rightarrow R_{56}(s)+\frac{L}{\gamma^2}\right]$. Here we define the microbunching gain function as $G(s) \equiv \left|g_k(s)/g_k^{(0)}(0)\right|$ and the gain spectral function at the exit of a lattice as $G_f(\lambda) \equiv G(s=s_f)$. We note that the impedance term in Eq. (6) is of our primary interest. With given impedance models [Eqs. (1-4)], we can estimate the microbunching gain through a beamline. Since the calculation is fast (compared with tracking simulation), it can be used to make quick estimation or optimize the microbunching gain development in a lattice design.

As of our second method, the particle tracking, we use ELEGANT [10] as our benchmarking tool to simulate the CSR-induced μBI in our example lattices. Detailed description of CSR tracking algorithms used in ELEGANT can be found in Ref. [11].

## SIMULATION RESULTS

To demonstrate how our semi-analytical simulation is used to analyze the microbunching gain for general linear lattices, we first take two 1.3 GeV recirculation arcs as our study examples. For the detailed description of the two example lattices, we refer the interested reader to Ref. [13]. Example 1 lattice is a 180° arc with large momentum compaction ($R_{56}$), which is a second-order achromat and globally isochronous with a large dispersion modulation across the entire arc. In contrast to the first example, Example 2 is again a 180° arc with however small momentum compaction. This arc is a second-order achromat designed to be a locally isochronous lattice within superperiods which insures that the bunch length is the same at phase homologous CSR emission sites.

For both lattices, the peak current is chosen to be 65.5 A, normalized emittance 0.3 μm, and uncorrelated energy spread $1.23\times10^{-5}$. CSR-induced microbunching gains for the two arcs are shown in Figs. 1 and 2. The two upper figures demonstrate the gain functions as a function of $s$ for two different modulation wavelengths with different combination of CSR models. One can see in Fig. 1 the shorter wavelengths enhance Landau damping through Eq. (7). The two bottom figures show the gain spectra $G_f(\lambda)$ at the exits of the lattices, from which one can obviously see the difference between them: Example 1 is vulnerable to CSR effect while Example 2 is not. To validate our semi-analytical results, we benchmark the two example lattices by ELEGANT, with which extensive convergence studies were performed [14], and both results show good agreement (see Figs. 1 and 2).

From bottom figure of Fig. 1, we find that the microbunching gain with the inclusion of both steady-state CSR and entrance transient effects is slightly lowered from the case of steady-state CSR alone. This is because the CSR impedances including entrance transient effect [Eq. (1) and (2)] become a bit reduced near a dipole entrance when the beam enters the bend. We also observe that, with the inclusion of all relevant CSR impedances, including exit transients, the microbunching gain increases up to 200 % compared with that of steady-state case. Note that, for this lattice, dipoles only occupy less than 5% of beamline length, so without optical compensation the CSR-drift transient can cause a significant effect. Yet with optical compensation, even with the same ratio of dipoles over the beamline, Example 2 is not subject to CSR-induced μBI (see Fig. 2). This highlights the impact of lattice design for recirculation arcs on microbunching gain [1]. We remind that, due to the extremely high gain of Example 1 lattice with inclusion of all relevant CSR impedances, those ELEGANT results are averaged over the initial amplitudes 0.01-0.04% and 70M macroparticles are used in the simulation.

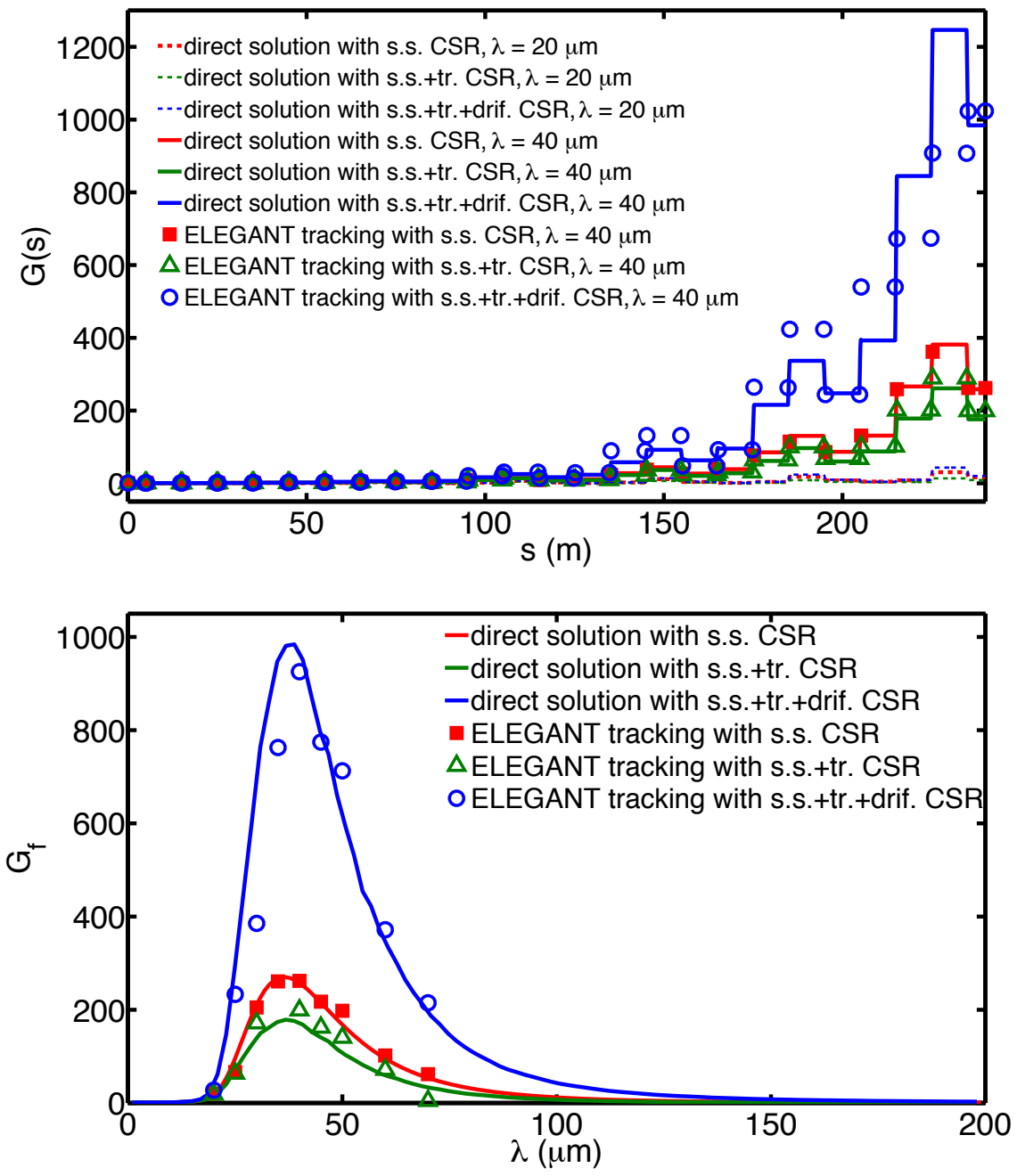

Figure 1: (Top) gain functions $G(s)$ as a function of $s$, where the dashed lines are for $\lambda$ = 20 mm and solid lines for $\lambda$ = 40 mm. (Bottom) gain spectra $G_f(\lambda)$ as a function of modulation wavelength. In ELEGANT simulation, the initial density modulation is set 0.05 % for steady-state (s.s.) case; 0.06 % for s.s. and entrance transient (tr.) case; 0.01-0.04 % for s.s., tr. and exit transient (drif.) case.

As the third example, we consider a medium-energy (750 MeV) arc compressor transporting a high-brightness electron beam with nearly no beam quality degradation [15]. Such bunch compressor arc is made up of total 9 combined-function dipoles. Those dipoles are wittingly designed with gradually reduced lengths in order to avoid the increasingly severe CSR effects among the downstream bending magnets. To compare our semi-analytical results with ELEGANT, we only keep the linear field components (dipole and quadrupole) and turn off sextupole fields, usually utilized for longitudinal phase space manipulation, in the combined-function dipoles in ELEGANT tracking. Further study of this compressor arc is in progress. For this example lattice: the peak current after compression is 405 A, normalized emittance 0.75 μm, uncorrelated energy spread $1.13\times10^{-5}$, and chirp -0.635 $m^{-1}$. The reader is referred to Ref. [15] for further information about this compressor arc design.

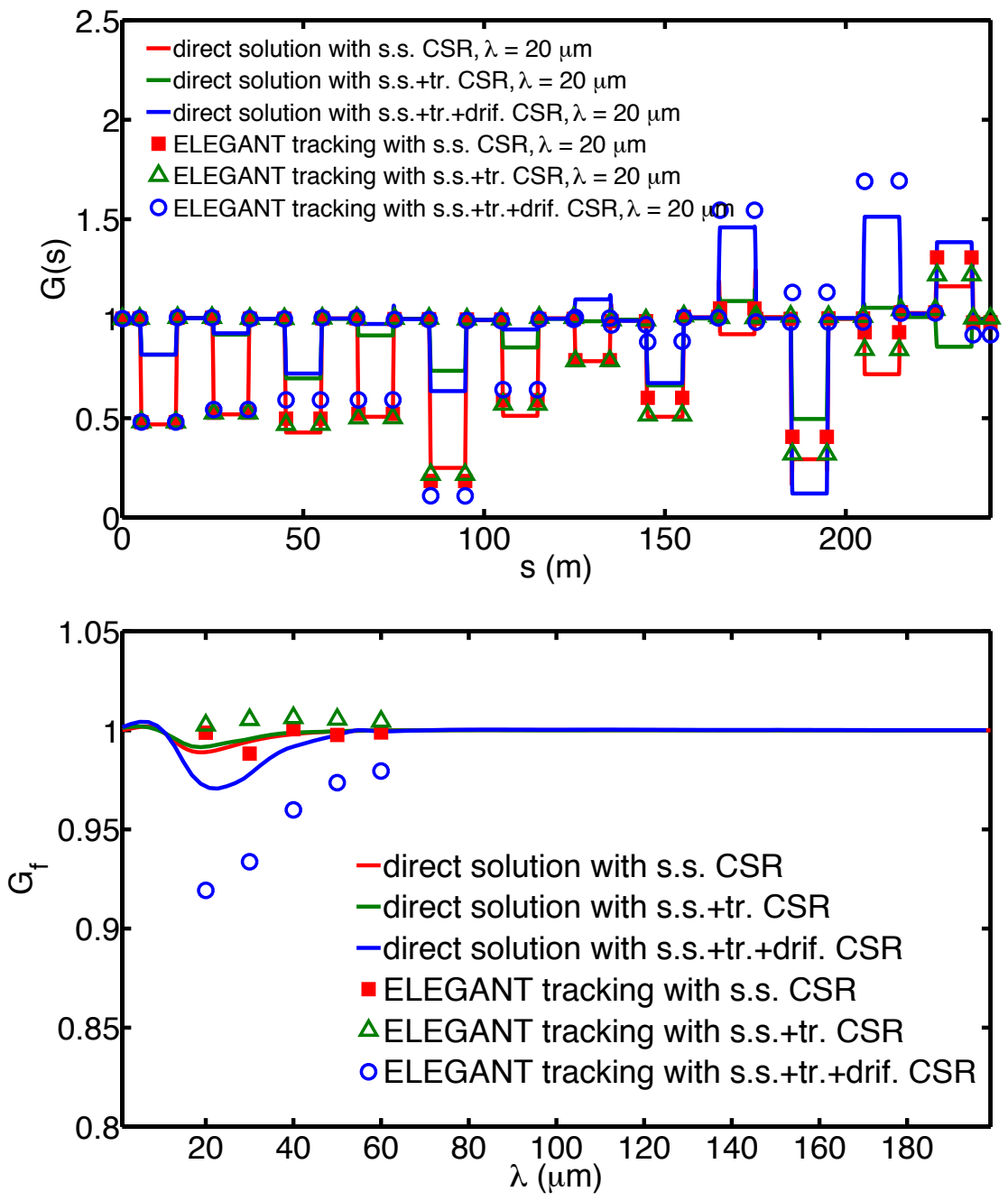

Figure 2: (Top) CSR gain functions $G(s)$; (bottom) gain spectra $G_f(\lambda)$ as a function of modulation wavelength. Initial density modulations are all set 0.8% in ELEGANT.

Figure 3 illustrates the gain functions $G(s)$ and the corresponding gain spectra at the exit of the compressor arc. The overall gain around the arc being less than 3 indicates that such artful design can make the transported beam almost refrain from CSR microbunching effect. We note that the gain peaks at wavelength around 4 mm, comparable to the design rms bunch length. This makes the coasting-beam assumption used in our Vlasov analysis no longer valid. In addition, for this parameter regime, the wall shielding effect should also be taken into account. Note that with the optical compensation carefully devised for this example, even with the inclusion of both entrance and exit transient effects, the microbunching gain increases up to only 8% (Vlasov solution) or 15% (ELEGANT) above the results of steady-state CSR case.

## SUMMARY

In this paper, we have extended our previous work [1] to further include more relevant CSR impedance models. Our results are benchmarked by ELEGANT and they achieve excellent agreement. The gain functions and spectra are shown to illustrate Landau damping effect and gain enhancement when including both steady-state and transient CSR effects (Figs. 1-3). For more detailed analyses of CSR gains with steady-state impedance and lattice impact on the microbunching amplification development, the interested reader is referred to Ref. [16].

By observing these gain spectra (Figs. 1-3) as well as the general features of the impedance models, we conclude that the microbunching gain with inclusion of only steady-state and entrance transient effects would give reduced values compared with that of steady-state case. However, the microbunching gain with inclusion of all CSR impedances [Eqs. (1-4)] would enhance over its steady-state gains. Finally we note Eqs. (1-3) assume ultrarelativistic beam. For those low beam energy case, e.g. a merger in an energy-recovery linac, simulating CSR-induced microbunching gain indeed requires more accurate impedance description, in which case our currently introduced models may not be valid.

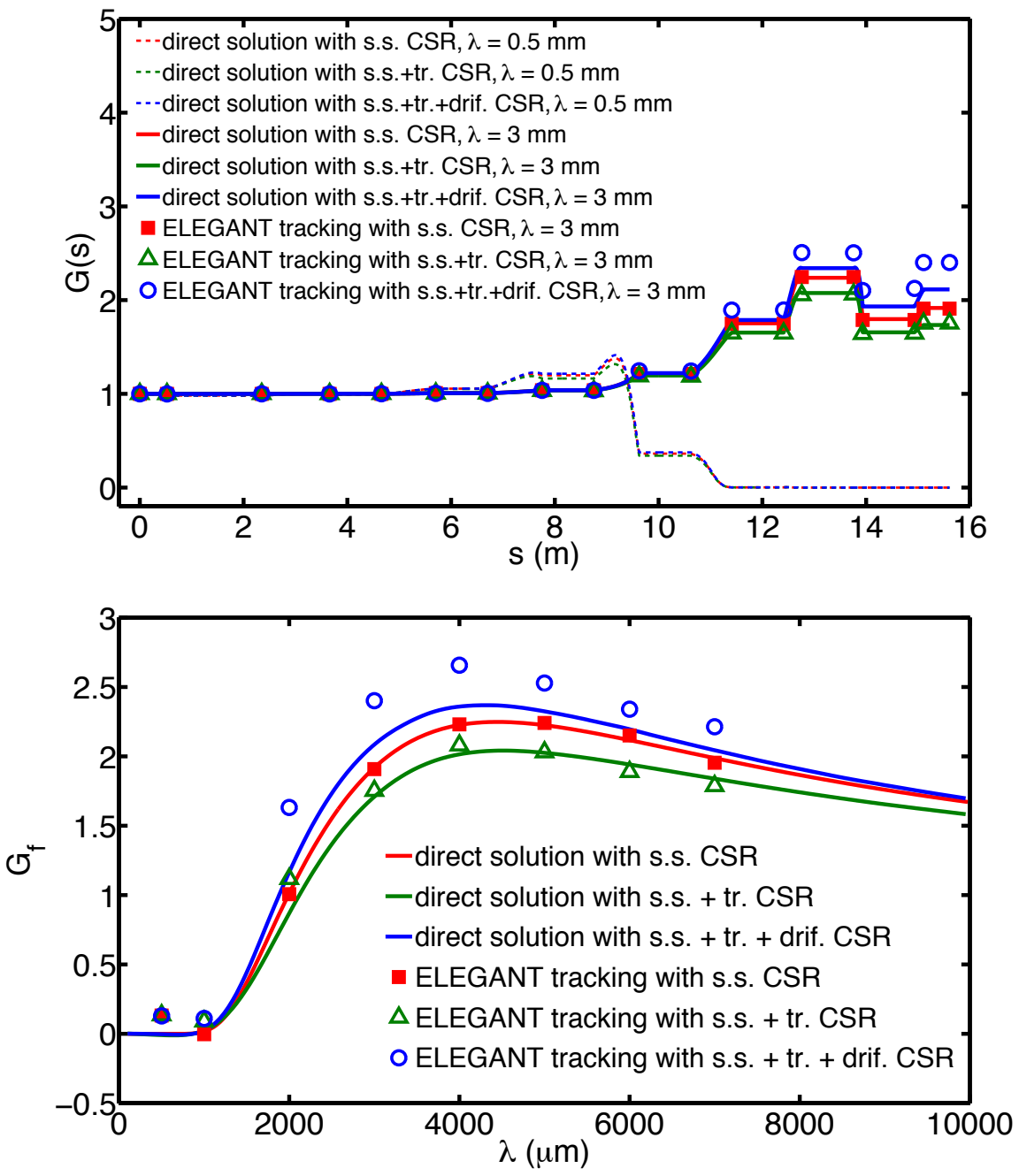

Figure 3: (Top) CSR gain functions $G(s)$, where the dashed lines are for $\lambda$ = 0.5 mm and solid lines for $\lambda$ = 3 mm. (Bottom) gain spectra $G_f(\lambda)$. Here in ELEGANT simulation the initial density modulations are all set 1 %.

## ACKNOWLEDGEMENTS

We thank Steve Benson for many stimulating and inspiring discussion.